# INTERNAL CONVERSION AND INTERSYSTEM CROSSING PATHWAYS IN UV EXCITED, ISOLATED URACILS AND THEIR IMPLICATIONS IN PREBIOTIC CHEMISTRY

Hui Yu, Jose A. Sanchez-Rodriguez, Susanne Ullrich*

Department of Physics and Astronomy, University of Georgia, Athens, GA 30602, USA

*corresponding author: ullrich@physast.uga.edu

M. Pollum, C. E. Crespo-Hernández

Department of Chemistry and Center for Chemical Dynamics, Case Western Reserve University, Cleveland, OH 44106, USA

Sebastian Mai, Philipp Marquetand, Leticia González

Institute of Theoretical Chemistry, University of Vienna, Währinger Str. 17, 1090 Vienna, Austria

**Abstract**

The photodynamic properties of molecules determine their ability to survive in harsh radiation environments. As such, the photostability of heterocyclic aromatic compounds to electromagnetic radiation is expected to have been one of the selection pressures influencing the prebiotic chemistry on early Earth. In the present study, the gas-phase photodynamics of uracil, 5-methyluracil (thymine) and 2-thiouracil—several heterocyclic compounds thought to be present during this era—are assessed in the context of their recently proposed intersystem crossing pathways that compete with internal conversion to the ground state. Specifically, time-resolved photoelectron spectroscopy measurements evidence femtosecond to picosecond timescales for relaxation of the singlet $^1\pi\pi^*$ and $^1n\pi^*$ states as well as for intersystem crossing to the triplet manifold. Trapping in the excited triplet state and intersystem crossing back to the ground state are investigated as potential factors contributing to the susceptibility of these molecules to ultraviolet photodamage.

**Introduction**

An important premise of the molecular origins of life is the availability of a large variety of organic heterocycles in the so-called "primordial soup" to supply the building blocks for genetic material.[1,2] This hypothesis is supported by the compositional analysis of meteorites,[3,4] prebiotic chemistry simulation experiments,[5,6,7,8,9,10] and the widespread use of non-canonical nucleobases across all domains of life.[2,11,12] These heterocycles can range from being extensively different from the canonical nucleobases, to containing just a single atom alteration, which retains the base pairing properties.[2,13] Two important examples of nucleobase modifications include thymine, which itself is a base modification of uracil (i.e., 5-methyluracil, see Figure 1), and 2-thiouracil, which has been found in transfer ribonucleic acids (tRNAs) from all domains of life. In particular, 2-thiouracil has been proposed as a prebiotic nucleobase with the ability to enhance the fidelity of nonenzymatic template-directed synthesis of RNA.[2,14] In their investigations, Zhang et al. found that thionation of the canonical nucleobases can increase both the polymerization rate and the copying accuracy of short RNA strands.[14] This observation led the authors to propose that 2-thiouracil could have played an important role in the self-replication processes of primordial RNA, particularly before the evolution of enzyme-catalyzed replication. Hence, the continued use of 2-thiouracil in tRNAs could be considered as a relic of the chemical origins of life.[14,15]

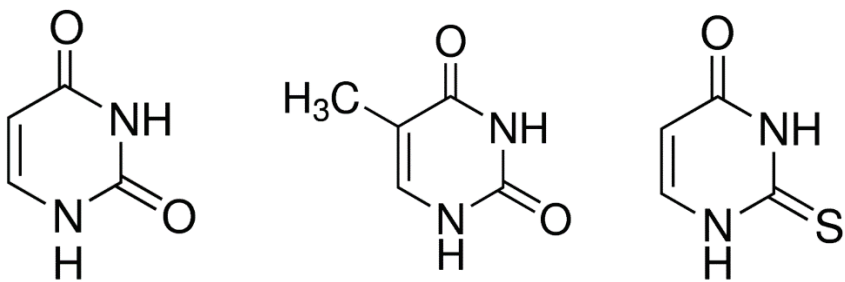

Fig. 1: Molecular structure of uracil (left), 5-methyluracil (thymine, middle), and 2-thiouracil (right) in their most stable tautomeric form which is observed under gas-phase, molecular beam conditions.

Besides facilitating replication fidelity and the transfer of information, another important requirement for prebiotic genetic material is the ability to retain chemical integrity under the harsh radiation environment on early Earth.[2] Photophysical pathways that link bright singlet excited states, populated through the absorption of an ultraviolet (UV) photon, back to the ground state, provide molecules with inherent photoprotective properties. Conical intersections (CI) between these states facilitate efficient internal conversion that can occur on ultrafast timescales. The canonical nucleobases are generally considered to be photostable under UV radiation due to their ultrafast excited-state deactivation to the ground state either directly or *via* intermediate singlet excited states.[16,17,18,19,20,21,22,23,24] This view has recently been challenged, with a number of studies reporting that a small fraction of the excited-state population might become trapped for prolonged times, *e.g.*, in dark singlet states or meta-stable triplet states.[25,26,27,28,29,30] For non-canonical nucleobases the extent of this trapping can vary considerably, which could have important consequences for their photostability. For example, recent experimental[31,32,33,34] and theoretical[35,36,37,38,39] works have shown that 2-thiouracil is efficiently trapped in a triplet excited state in nearly unity yield following UV excitation in solution. Motivated by these findings, the present work investigates the excited-state dynamics of uracil derivatives in the gas phase using time-resolved photoelectron spectroscopy (TR-PES) supported by quantum-chemical calculations in order to better understand the interplay between ultrafast internal conversion to the ground state and the intersystem crossing (ISC) pathway. Understanding how the electronic relaxation mechanisms of the nucleobases are altered by slight modifications to their chemical structure may be paramount to tracing the lineage of these molecules of life.

Despite the substantial amount of time-resolved work on 2-thiouracil in the condensed phase,[31,32,33,34] to the best of our knowledge no gas-phase time-resolved photoionization-based measurements, neither TR-IY (time-resolved ion yield) nor TR-PES, have yet been reported in the literature for this molecule. A main objective of the present study is to present the first TR-PES spectrum of 2-thiouracil and, in conjunction with the TR-PES of uracil and thymine, to evaluate the major contributing steps to the photophysical deactivation of uracil derivatives following photoexcitation to their $S_2$ ($\pi\pi^*$) state. Specifically, the following pathways are scrutinized: (1) trapping in the $S_2$ minimum and barrier crossing *versus* direct paths toward the $^1\pi\pi^*/S_0$ or the $^1\pi\pi^*/^1n\pi^*$ CIs. (2) ground state repopulation *via* the $^1\pi\pi^*/S_0$ CI *versus* deactivation to and trapping on the $S_1$ ($^1n\pi^*$) surface, (3) the fate of the $S_1$ ($^1n\pi^*$) state population, which has been proposed to deactivate to both the ground state and the triplet excited states. TR-PES provides spectroscopic information that, *via* known ionization potentials and correlations, can be used to identify the orbital configurations of the excited states. As such, these spectroscopic capabilities combined with an extended experimental time window, allow for further investigation of any long-lived states and the dynamics potentially associated with ISC. In terms of photostability, ultrafast photophysical processes that lead to repopulation of the ground state provide protective properties under harmful UV irradiation. However, understanding the slower relaxation dynamics is equally important as molecules that remain trapped in electronically excited states for extended periods of time are subject to increased photodamage under continued UV irradiation.

**Experimental Setup**

The experimental setup used for the present TR-PES measurements includes a gaseous molecular beam source, a magnetic bottle photoelectron spectrometer, a linear time-of-flight mass spectrometer, and a femtosecond laser system with UV conversion capabilities that have been described previously.[19,40,41,42,43]

Uracil, thymine and 2-thiouracil (Sigma Aldrich, ≥98%) were placed in a quartz sample holder which is located inside the nozzle just before the pinhole and heated to 220°C, 175°C, and 210°C, respectively. A continuous molecular beam of sample vapor was carried into the source vacuum chamber by a helium backing gas and doubly skimmed to accommodate differential pumping towards the ultrahigh vacuum chamber that houses the photoelectron energy analyzer and mass spectrometer. The molecular beam is intersected by a femtosecond UV pump pulse and a time-delayed probe pulse that are focused by 50 cm lenses and spatially overlapped at a small angle.

Pump pulses centered at 260 nm and 290 nm were generated from a Traveling-wave Optical Parametric Amplifier (TOPAS-C) and kept at 1.5-2 μJ / pulse to avoid two-photon excitation. Probe pulses were produced in a second OPA (Coherent OPERA) and set to 295 nm for uracil and thymine and 330 nm for 2-thiouracil, just below the onset of their absorption spectra, to avoid unwanted probe-pump signals. This wavelength choice imposes the requirement of a two-photon process for ionization and hence higher probe pulse energies of 12-15 μJ / pulse were employed. For uracil and thymine, short range TR-PES scans were recorded between -1 ps < Δt < 6 ps with a 25 fs step size and long range scans were taken between -1 ps < Δt < 600 ps with unequal step size. The 2-thiouracil scan ranges were -1 ps < Δt < 4 ps and -1 ps < Δt < 200 ps for the short and long TR-PES scans, respectively.

Timing calibrations of all TR-PES data are based on 1,3-butadiene measurements which yield a Gaussian cross-correlation function with typically ~200 fs full width at half maximum (FWHM) and define the position of zero pump-probe delay. TR-PES spectra are energy calibrated with 1,3-butadiene using known ionic state potentials.[44] Variant repelling and retarding voltages were applied in the ionization region to collect photoelectrons of certain kinetic energies as outlined in Refs. 41 and 42.

Additional time-resolved ion yield (TR-IY) spectra (not presented here) were recorded to confirm the dominance of the parent compound in the mass spectrum under the chosen molecular beam conditions. The mass spectrum of 2-thiouracil is provided in the Supplementary Information since no prior molecular-beam based TR-IY (or TR-PES) measurements of this compound are available in the literature. Our experimental technique does not allow us to distinguish between tautomers. However, *ab initio* studies and gas-phase spectroscopies have identified the oxo-thione form (Fig. 1 (right)) as the most stable and dominant tautomer of 2-thiouracil.[45,46,47] Furthermore, no thermal decomposition of the sample in the heated nozzle was observed by Ref. 46. Phototaumerization has been shown to occur on long (minutes) time scales and is unlikely of any relevance to the present study.[48] Similar studies performed on uracil and thymine identify the diketo form as their dominant tautomer (Fig. 1 (left) and (middle), respectively).[49,50] We therefore assume that our TR-PES spectra are associated with these tautomers as is illustrated in Fig. 1.

**Computational Details**

For 2-thiouracil, quantum chemical calculations were performed to rationalize the experimental shift in the ionization potential. Based on the geometries from the linear interpolation scans in Ref. 36, we performed MS-CASPT2(n,9)/ano-rcc-vtzp calculations (multi-state complete active space perturbation theory of second order, with n electrons in 9 active orbitals, n being 12 for neutral states and 11 for ionic ones).[51,52] 4 singlets, 5 doublets and 3 triplets were calculated. Furthermore, a minimum energy crossing between $T_1$ and $S_0$ was optimized[53] with MS-CASPT2(12,9) with the smaller cc-pVDZ basis set. All calculations were performed with MOLCAS 8.0.[54]

**Results and Discussion**

As structurally similar molecules, the UV absorption spectra of uracil and thymine bear close resemblance, characterized by a first absorption band spanning 300-220 nm (4.1-5.6 eV) which evolves into a second broad feature around 220-180 nm (5.6-5.9 eV) or 220-163 nm (5.6-7.6 eV), respectively.[55,56] Thus, both uracil and thymine were excited at 260 nm and probed by two-photon absorption at 295 nm to produce their TR-PES shown in Figure 2. Heavy atom substitution manifests itself as a significant redshift of the 2-thiouracil absorption spectrum compared to uracil, with the first absorption band spanning 320-240 nm (3.9-5.2 eV) and a very weak absorption extending to approximately 350 nm (3.5 eV).[48] Therefore, the TR-PES of 2-thiouracil were collected using a 290 nm pump and 330 nm two-photon probe. In each case, the pump wavelengths were chosen to populate the $S_2$ band of $^1\pi\pi^*$ character[56,57,58] with a limited degree of vibrational excitation (i.e. on the rising edge of the lowest energy absorption band) in order to assess deactivation pathways involving the lowest singlet and triplet electronic excited states.

The TR-PES spectra of uracil, thymine, and 2-thiouracil in Figure 2 are displayed from top to bottom, respectively, arranged in the following manner. Column 1 shows the recorded 2d TR-PES signal over the first few picoseconds. Global analysis based on simultaneous fitting of the spectra and associated dynamics yielded the individual contributions (channels) to the 2d data, which are displayed in columns 3-5, and their summation (*i.e.* total fit), presented in column 2 for comparison to the recorded data in column 1. It should be noted that in the kinetic modeling of the data, the spectrum of each channel is assumed to be independent of time, *i.e.* all signal amplitudes composing the spectrum follow the same exponential decay. This assumption poses limitations on the data analysis, in particular, in cases where the cationic ionization potentials vary significantly along the relaxation path. However, full *ab initio* simulation of the experimental TR-PES spectra is considered to be beyond the scope of this work.

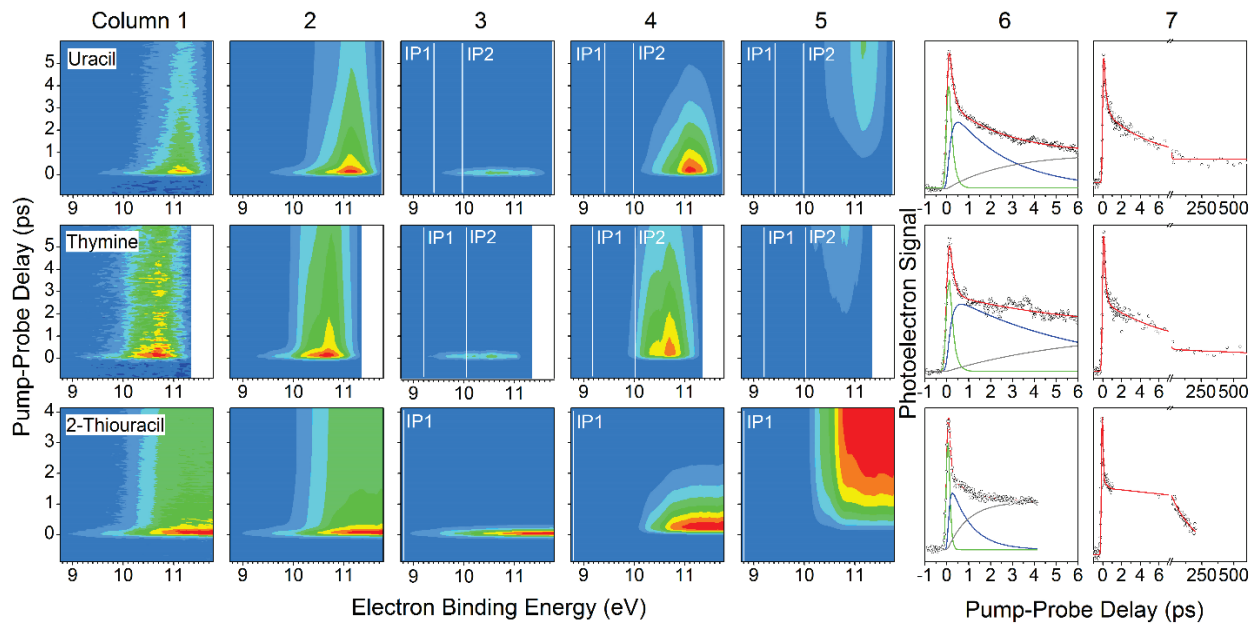

Fig. 2: TR-PES of uracil (top), thymine (middle) and 2-thiouracil (bottom) following excitation to their $S_2$ ($\pi\pi^*$) state. Pump wavelengths of 260 nm for uracil and thymine and 290 nm for 2-thiouracil were employed. The two-dimensional TR-PES spectrum, total fit, and individual contributions obtained *via* global analysis are shown for each molecule (displayed from left to right), as well as the corresponding integrated time traces: signal (open circles), fit (red solid line), and individual contributions (green, blue, and grey solid lines). Vertical ionization potentials from He(I) photoelectron spectroscopy are indicated in columns 3-6 as white lines with labels IP1 and IP2. The dynamics of the long-lived channel were extracted from a long-range TR-PES scan and the corresponding time trace of the total integrated signal and fit are shown in the graphs at the far right.

All 2d data is plotted as a function of electron binding energy *versus* pump-probe delay, with the electron binding energy being readily calculated as the total photon energy minus the measured photoelectron kinetic energy, *i.e.*:

$$E_{bind}(t) = E_{photon} - E_{el.kin.}(t). \qquad (1)$$

The corresponding time traces are obtained by integrating the 2d data over all electron binding energies. They are shown in column 6 including the overall signal (integrated column 1, circles), total fit (integrated column 2, red line) and individual contributions (integrated columns 3-5, green, blue and grey lines). Time traces of separate, long-range TR-PES scans used to extract the long-lived time constants for each molecule are displayed in column 7 including the total fit (red line) to the integrated signal (circles). Fitting functions are based on sequential exponential decays convoluted by the Gaussian instrument response function. All time constants are collated in Table 1. Specifically, the time constant $\tau_1$ corresponds to the decay of the TR-PES signal in column 3 and the green exponential decay trace in columns 6; time constant $\tau_2$ describes the decay in column 4 and the blue trace in column 6; and time constant $\tau_3$ is associated with the decay in column 5 and the grey trace in column 6.

Table 1: Time constants for sequential decay dynamics in uracil, thymine and 2-thiouracil as extracted from the TR-PES spectra presented in Figure 2. An error of ~20% is estimated based on the fitting statistics.

| Molecule | $\tau_1$ | $\tau_2$ | $\tau_3$ |
| --- | --- | --- | --- |
| Uracil | 170 fs | 2.35 ps | >1 ns |
| Thymine | 175 fs | 6.13 ps | >1 ns |
| 2-Thiouracil | <100 fs | 775 fs | ~203 ps |

The photoelectron spectra are analyzed based on ionization correlations between the neutral excited states and lowest cationic states and reported or calculated ionization potentials obtained from *ab initio* calculations and spectroscopic measurements as given in the discussion below. In particular, we calculate the total binding energy at a geometry R as the sum of the corresponding adiabatic IP at R and the vibrational energy gain of the neutral excited state during relaxation from the FC region to R:

$$E_{bind.}(R) = \underbrace{E_{ion}(R) - E_{S_0}(FC)}_{adiab.\ IP(R)} + \underbrace{E_{bright}(FC) - E_{neutral}(R)}_{E_{vib.\ gain}}. \qquad (2)$$

This theoretical value allows for a correlation of the experimental spectral shifts with the theoretically predicted state-to-state transitions.

In the following sections, we initially focus our discussion on uracil and thymine due to the recent attention they have received in various experimental[59,60,61,62,63,64,65,66,67,68] and theoretical[24,27,69,70,71,72,73,74,75,76,77] studies. This is followed by 2-thiouracil, which in contrast has received considerably less attention, with limited studies from a theoretical standpoint[35,36] and currently no photoionization-based spectroscopic studies in the gas phase.

**Uracil**

High-resolution resonance enhanced multi-photon ionization spectra of uracil are broad with an onset of 36600 cm$^{-1}$ (4.538 eV).[78] This agrees very well with the S$_2$ ($^1\pi\pi^*$) minimum energy of 4.53 eV calculated by Yamazaki et al.[76] Based on these values, initial photoexcitation of uracil at 260 nm to the S$_2$ ($^1\pi\pi^*$) state results in an estimated excess energy of 0.23 eV. This state preferentially ionizes into the first ionization potential (IP1), which was located by high-resolution mass-analyzed threshold ionization (MATI) to be at 9.3411 eV[79] and by conventional He(I) photoelectron spectroscopy to be at 9.60 ($\pi^{-1}$); the latter study also reports IP2 as 10.13 eV (n$^{-1}$).[80] Both IPs are indicated in the 2d spectra (Fig. 2, column 3-5) as lines with labels IP1 and IP2. These ionization energies generally agree with the theoretical estimates of 9.41 eV and 10.11 eV by Matsika et al.[77] and 9.56 eV and 10.28 eV by Martinez and coworkers,[70] but are also predicted to change significantly along the relaxation path. Of relevance to our analysis, Matsika et al. compute IP1 as 10.42 eV at the geometry of the S$_2$ minimum, which is a ~1 eV shift compared to the Franck-Condon region. Assuming Δv = 0 for the ionization process (conservation of vibrational excitation during ionization), based on the reported IP1 at the S$_2$ minimum and the vibrational excess energy of 0.23 eV, we expect that ionization from the S$_2$ minimum will yield a photoelectron band centered at 10.65 eV which is indeed observed here (see Figure 2, column 3). Based on this accurate agreement, together with the fact that S$_2$ is considered the only optically accessible state at the employed excitation wavelength, we assign the photoelectron spectra in Fig. 2 column 3 around zero pump-probe delay to the S$_2$ ($^1\pi\pi^*$) state. According to this photoelectron signal, relaxation to the S$_2$ minimum occurs on an ultrafast timescale, within the time resolution of the experiment.

The fate of the S$_2$ ($^1\pi\pi^*$) population can be determined by comparison of the photoelectron spectra in columns 3 and 4 of Fig. 2. Most theoretical studies predict internal conversion to the S$_1$ ($^1n\pi^*$) state as the next step in the deactivation process,[20,21,22,24,81,82,83] which involves a switch in electronic state character that is accompanied by changes in ionization preferences (i.e. S$_2$ ($^1\pi\pi^*$) → IP1 but S$_1$ ($^1n\pi^*$) → IP2). However, for ionization from the respective excited-state minima these IPs are almost degenerate (10.42 vs. 10.48 eV).[77] As uracil decays to the lower lying S$_1$ ($^1n\pi^*$) minimum, additional vibrational energy is gained amounting to the difference in electronic state energies (~0.38 eV based on Ref. 76). This is reflected in the experimental photoelectron spectra by a shift of the signal in column 4 to approximately 11 eV. We can therefore associate the time constant $\tau_1$ (~170fs) in Table 1, row 1 with $S_2$ ($^1\pi\pi^*$) to $S_1$ ($^1n\pi^*$) relaxation as well as direct $S_2$ ($^1\pi\pi^*$) to $S_0$ deactivation. These processes occur simultaneously and are indistinguishable in our TR-PES spectra. Our data does not support extended trapping of population in the $S_2$ ($^1\pi\pi^*$) minimum for picosecond timescales as previously predicted by Refs. 70 and 71.

The second channel, associated with the $S_1$ ($^1n\pi^*$) state (Figure 2, column 4), deactivates to populate a third state (Figure 2, column 5) on an intermediate picosecond timescale (Table 1, row 2, $\tau_2$), while internal conversion back to the electronic ground state is likely to also contribute to the signal decay in column 4. Although adequate description of the slow dynamics necessitates a third, long-lived (nanosecond) component in the fit, no obvious spectral differences are observed between columns 4 and 5. The signal in column 5 is therefore suggestive of either extended trapping on the $S_1$ ($^1n\pi^*$) state or of a photophysical relaxation process where energetic shifts due to vibrational and electronic changes coincidentally cancel each other. At this point, a pathway based on ISC recently proposed by Etinski *et al.*[83] and Richter *et al.*[27] requires consideration. According to their proposal, the $S_2$ ($^1\pi\pi^*$) to $S_1$ ($^1n\pi^*$) internal conversion occurs on ultrafast timescales followed by ISC to the $T_1$ ($^3\pi\pi^*$) state. Unfortunately, no *ab initio* calculations pertaining to photoionization of intermediates along this pathway were pursued by the authors. Our discussion is therefore limited to the expected vibrational shifts due to electronic relaxation. According to the adiabatic excitation energies given by Etinski *et al.*,[83] the $T_1$ ($^3\pi\pi^*$) minimum is located ~0.9 eV below the $S_1$ ($^1n\pi^*$) minimum and consequently the system would acquire this amount of energy as vibrational excitation during this relaxation step (column 4 to 5). In analogy to the $S_2$ ($^1\pi\pi^*$) state, preferential ionization of the $T_1$ ($^3\pi\pi^*$) state into IP1 is assumed but now with a significant shift toward higher electron binding energy due to the additional vibrational excitation (see column 5 in comparison to column 3). This analysis, at least qualitatively, lends support to the theoretical ISC pathway with the $S_1$ ($^1n\pi^*$) state acting as the doorway for triplet state population. *Ab initio* modeling of the TR-PES data is necessary and may be able to explain the lack of spectral shift during the ISC process.

Overall, there is generally good agreement between our time constants (Table 1) and those in the existing literature;[22,59,60,61,66,77] however, there is some discrepancy in assignment of the observed lifetimes. The ns decay reported here confirms the observation of a long-lived contribution in recent measurements by Matsika and co-workers[77] and Ligare et al.[67] Matsika et al.'s TR-IY measurements were analyzed for their strong field dissociative ionization pattern and dynamical evolution of major mass peaks. The authors assigned the TR-IY signals observed at >10 ps in the m/z = 69 fragment, but not the parent, to be due to $S_1$ ($^1n\pi^*$) ionization because of its higher tendency for dissociative ionization. Consequently, their fs and few ps lifetimes were associated with processes in the $S_2$ ($^1\pi\pi^*$) state. Photon impact mass spectrometry studies by Jochims and co-workers[84] have reported an appearance energy of 10.95 eV for the onset of cationic dissociation into the m/z = 69 fragment. Ionization from the $S_2$ ($^1\pi\pi^*$) minimum into IP1 falls mostly below this onset indicating that cationic fragmentation into m/z = 69 should be minimal. In comparison, ionization of the $S_1$ ($^1n\pi^*$) minimum correlates with IP2, with similar ionization energy as IP1, but with significant vibrational excitation acquired during the $S_2 \rightarrow S_1$ internal conversion process. In this latter case, the cationic state is accessed at an overall higher energy. Therefore, the TR-PES measurements presented in Fig. 2 question the assignments by Matsika et al. given that the intermediate picosecond and long-lived nanosecond channels both ionize into a similar electron binding energy range which falls right at the fragment appearance energy. On the other hand, Ligare et al. assigned the long-lived excited state observed in their double-resonance and nanosecond pump-probe experiments to the $T_1$ ($^3\pi\pi^*$) state based on its N-H stretching frequency, and it yielded a lifetime of several tens of nanoseconds. This is in agreement with our current assignment of the long-lived TR-PES signal in Fig. 2 column 5 to be that of the $T_1$ ($^3\pi\pi^*$) state.

**Thymine**

The thymine TR-PES data (Fig. 2, row 2) is analyzed analogous to the results for uracil and the extracted time constants are listed in Table 1. The onset of the experimental thymine excitation spectrum is located at 36300 cm$^{-1}$ (4.501 eV)[78] and calculated energies of various $S_0$, $S_1$ and $S_2$ geometries as well as the corresponding ionization energies can be found in Ref. 70.

The two lowest vertical ionization potentials are placed at 9.20 ($\pi^{-1}$) and 10.05 eV ($n^{-1}$) by conventional He(I) photoelectron spectroscopy,[80] and ionization correlations for thymine are similar to uracil, *i.e.* $S_2$ ($^1\pi\pi^*$) → IP1 and $S_1$ ($^1n\pi^*$) → IP2.[70] Initial excitation populates the $S_2$ ($^1\pi\pi^*$) state with an excess energy of 0.268 eV, which relaxes on ultrafast (~170 fs) timescales accompanied with a spectral shift from the 9.5-11 eV electron binding energy region to the 10-11.3 eV region (Fig. 2, column 3 *vs.* 4), again coinciding with the calculated IP1 and IP2 at the $S_2$ and $S_1$ minimum energy geometries, respectively.[70] In analogy to uracil, the underlying photophysical processes are associated with "instantaneous" evolution to the $S_2$ minimum followed by internal conversion either back to $S_0$ (direct pathway) or to the $S_1$ ($^1n\pi^*$) state. The photoelectron signal in the second channel (Fig. 2, column 4) decays on a picosecond timescale but in comparison to uracil this time constant increases from 2.35 to 6.13 ps. According to dynamics simulations,[23,24,70,73,75] access to the $^1\pi\pi^*/S_0$ CI either directly or *via* re-crossing from the $S_1$ ($^1n\pi^*$) state proceeds along a relaxation coordinate that involves motion of the $CH_3$ group out of the molecular plane, a process that is faster in uracil as only the light H-atom has to move. The $S_1$ ($^1n\pi^*$) again provides access to the long-lived (ns) triplet manifold, $T_1$ ($^3\pi\pi^*$), which is one of three ISC pathways predicted by Ref. 85.

In comparison to the existing literature, there is good agreement with previously reported time constants.[59,60,62,63,64,65,86] The ns decay component of thymine was also observed by Schultz and co-workers,[62,63,64,65] Kim *et al.*,[86] and Ligare *et al.*[67] Only the latter study offers a spectroscopic assignment to the $T_1$ ($^3\pi\pi^*$) based on double resonance measurements of the excited-state N-H stretch frequency and determines its lifetime as a few hundred nanoseconds. These observations for thymine further support the overall decay mechanism put forward above for uracil: $S_2$ ($^1\pi\pi^*$) → $S_1$ ($^1n\pi^*$) → $T_1$ ($^3\pi\pi^*$).

**2-Thiouracil**

At first glance the 2-thiouracil TR-PES data may appear similar to that of uracil and thymine, however, the ionization correlations and consequently data interpretation as discussed below are different. He(I) photoelectron spectra[87] and *ab initio* calculations[39] of 2-thiouracil associate the lowest IP (*i.e.* IP1) at 8.8 eV with two close-lying ionic states, a π-hole and a n-hole (both localized on the sulfur atom); therefore, electronically excited states of both ππ* and nπ* character are expected to preferentially ionize into IP1. Using the geometry data from Ref. 36, we computed the energies of the two lowest ionic states to assess the variation of IP1 along the relaxation path. Figure 3 displays the energetics along the relaxation path for the involved neutral states as well as the ionic states corresponding to IP1. In order to allow for a direct comparison with the TR-PES data, the figure also provides the total binding energy depending on molecular geometry, calculated according to equation (2). Note that for 2-thiouracil we use the average energy of $D_0$ and $D_1$ (the two states contributing to IP1), because for most geometries ionization to both states is equally probable. The total binding energy predicts the location of the photoelectron band in the TR-PES spectrum when plotted in terms of electron binding energy. According to theory, the $S_2$ minimum provides almost barrierless access to the $^1\pi\pi^*/^1n\pi^*$ CI and should facilitate ultrafast depopulation of the $S_2$ ($^1\pi\pi^*$) state.[36] The short-lived channel (Figure 2, column 3) extracted from the TR-PES spectrum is consistent with such a process. Specifically, the onset of the TR-PES signal occurs at approximately 9 eV, in line with the computed total binding energy in the FC region (8.9 eV, Figure 3). However, the signal level increases significantly toward higher electron binding energies, indicating ionization from the $S_2$ ($^1\pi\pi^*$) minimum (total binding energy of 9.7 eV) and possibly from a subsequently populated lower excited state. We therefore assign the channel in column 3 to photoionization of the initially excited $S_2$ ($^1\pi\pi^*$) state which decays rapidly within <100 fs. The most prominent spectral changes in the TR-PES are observed between columns 3 and 4 with the onset of the photoelectron band in the latter channel shifting ~1 eV higher in electron binding energy and a maximum signal level just above 11 eV. This indicates an increase in IP1 due to destabilization of the ionic states and vibrational energy gain, in line with *ab initio* predictions for the decay of the $S_2$ ($^1\pi\pi^*$) state to populate the $S_1$ ($^1n\pi^*$) state (see Figure 3). The signal in column 4 is therefore associated with ionization of the $S_1$ ($^1n\pi^*$) state which decays within 775 fs to populate the triplet manifold. As expected, the photoelectron band remains similar in column 5, given that only slight changes in geometry or energy occur during the $S_1$ ($^1n\pi^*$) to $T_2$ ($^3n\pi^*/^3\pi\pi^*$) ISC and the subsequent internal conversion to the $T_1$ ($^3\pi\pi^*$) minimum (see Figure 3). This is in agreement with the theoretical predictions of Ruckenbauer *et al.*,[39] who reported that the photoelectron spectra of 2-thiouracil are very similar for ionization from $S_1$ ($^1n\pi^*$) and $T_1$ ($^3\pi\pi^*$). In summary, our TR-PES spectra support the $S_2$ ($^1\pi\pi^*$) → $S_1$ ($^1n\pi^*$) → $T_2$ ($^3n\pi^*/^3\pi\pi^*$) → $T_1$ ($^3\pi\pi^*$) pathway proposed in Ref. 35 and 36. It should be noted however that our TR-PES spectra do not necessarily exclude the alternative pathway proposed by Borin *et al.*[35] wherein ISC occurs between $S_2$ ($^1\pi\pi^*$) and $T_2$ ($^3n\pi^*$), since our experimental technique cannot unambiguously distinguish between the $^1n\pi^*$, $^3n\pi^*$, and $^3\pi\pi^*$ states involved. However, the $S_2$ ($^1\pi\pi^*$) → $T_2$ ($^3n\pi^*$) → $T_1$ ($^3\pi\pi^*$) pathway is considered unlikely since it would have to outcompete the ultrafast and efficient $S_2$ ($^1\pi\pi^*$) to $S_1$ ($^1n\pi^*$) internal conversion, which was found to occur on a sub-100 fs timescale.

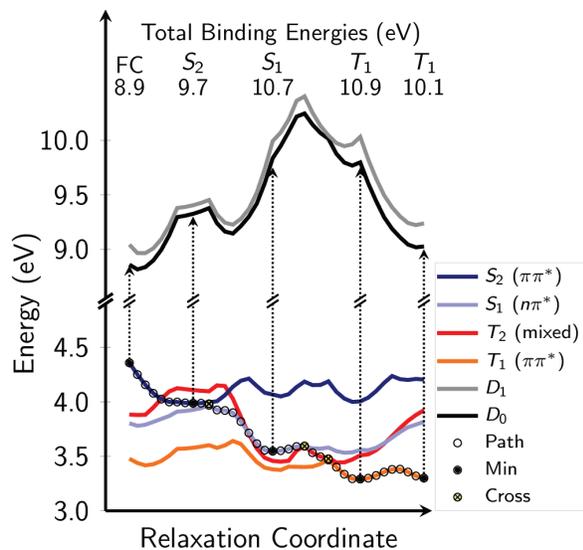

Fig. 3: Potential energy scan along the pathway connecting the $S_2$ ($^1\pi\pi^*$) Franck-Condon (FC) region to the $T_1$ ($^3\pi\pi^*$) state (Paths II and III from Ref. 36), using the MS-CASPT2/ano-rcc-vtzp level of theory. On the top, the total binding energies are listed for the respective excited-state minima. For a given geometry (R), the total binding energy includes the ionic state energy and vibrational energy gain of the neutral excited state during the relaxation from the FC region to that particular geometry along the pathway.

Finally, the third time constant for 2-thiouracil was found to be 203 ps, however, this value may only be approximated due to the limited time range scanned in this study (~200 ps, see Figure 2). We assign this constant to ISC from the $T_1$ ($^3\pi\pi^*$) state back to the electronic ground state. Notably, while the rate of triplet state decay for uracil and thymine increases a single order of magnitude in going from the condensed[88] to the gas phase,[67] this rate of triplet state decay for 2-thiouracil is at least two orders of magnitude faster than that measured in solution (70 ns in acetonitrile[31,33]). The triplet decay rate of other thiopyrimidine derivatives has previously been observed to increase with decreasing solvent polarity,[89] but this is the first time the rate of triplet state decay of a thiopyrimidine has been measured in the gas phase. Thus, in order to further scrutinize our assignment, we optimized a $T_1/S_0$ crossing point (see Fig. 4 for the geometry and energies), at the MS-CASPT2(12,9)/cc-pVDZ level of theory (as used in Ref. 36). An interpolation scan from the pyramidalized $T_1$ ($^3\pi\pi^*$) minimum[36] (whose geometry is similar to that at the crossing point) shows only a small barrier of 0.2 eV to reach the crossing point with $S_0$ and a SOC of about 130 cm$^{-1}$ at this point. These values are consistent with a relatively rapid deactivation of the $T_1$ ($^3\pi\pi^*$) state back to the ground state observed in the gas phase measurements. For comparison, identical calculations performed for uracil (see Supplemental Information) reveal a barrier of 0.4 eV and SOC of 1-2 cm$^{-1}$ and for thymine values of 0.13 eV and 2 cm$^{-1}$ have been reported by Ref. 85 although at a slightly different level of theory. The difference in SOC easily accounts for a factor of 3-4 orders of magnitude in the observed lifetimes between 2-thiouracil on the one hand and the natural nucleobases uracil and thymine on the other hand.

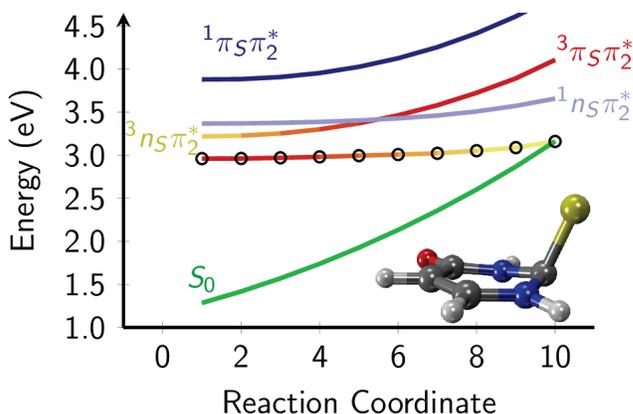

Fig. 4: Linear interpolation scan from the pyramidalized[36] $T_1$ minimum of 2-thiouracil to a $T_1/S_0$ crossing at the MS-CASPT2(12,9)/cc-pVDZ level of theory. Color indicates the state character, which changes for the $T_1$ surface (marked with rings) from $^3\pi\pi^*$ (red) to $^3n\pi^*$ (gold) along the reaction coordinate. The geometry at the crossing point is shown in the lower right corner and the coordinates are listed in the Supplemental Information. A similar linear interpolation scan for uracil is also provided in the Supplemental Information.

**Conclusions**

The excited state dynamics of substituted uracils following photoexcitation to their bright $S_2$ ($^1\pi\pi^*$) state have been investigated using TR-PES. All three molecules display a multi-exponential decay with three distinct time constants: one ultrafast, one intermediate (hundreds of fs to few ps), and one long-lived (hundreds of ps to ns). The associated relaxation pathways are summarized in the schematic in Figure 5.

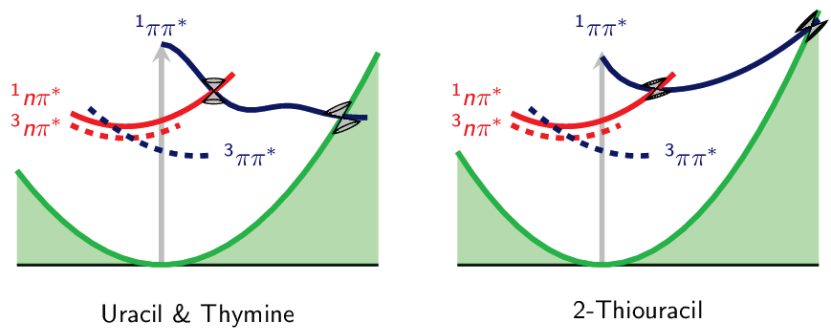

Fig. 5: Schematic of the relaxation pathways in uracil, thymine and 2-thiouracil following excitation to their $S_2$ ($^1\pi\pi^*$) state.

In the case of uracil and thymine, internal conversion from their $S_2$ ($^1\pi\pi^*$) state either back to $S_0$ or the lower lying $S_1$ ($^1n\pi^*$) state occurs within <200 fs. Population converting to the $S_1$ ($^1n\pi^*$) state becomes trapped, giving rise to lifetimes of 2.35 or 6.13 ps, respectively. This trapping allows a fraction of the population to access the triplet manifold, where it remains for nanoseconds.

Consistent with the heavy atom effect expected for sulfur substitution, relaxation times of the modified nucleobase 2-thiouracil are considerably faster, with <100 fs internal conversion to the $S_1$ ($^1n\pi^*$) state and subsequent population of the triplet state within 775 fs. *Ab initio* calculations of the neutral and ionic states were performed in order to interpret the TR-PES spectra. The quantum yield for triplet state population cannot be extracted from the TR-PES spectrum, however, based on *ab initio* calculations,[35,36] no ground state repopulation out of the $S_1$ ($^1n\pi^*$) should occur. This is also consistent with the femtosecond ISC and near unity triplet yields previously measured for 2-thiouracil in solution.[32] ISC from $T_1$ ($^3\pi\pi^*$) back to the ground state is also very efficient, with a triplet decay lifetime of only about 203 ps obtained from the TR-PES measurement.

In terms of prebiotic chemistry and the general idea of natural selection of photostable molecules as the building blocks for life, the present study suggests that the highly-efficient triplet-state population of 2-thiouracil is not irreconcilable with prebiotic survival, if the triplet population can also efficiently deactivate back to the ground state, as reported herein for a gas phase environment. As for uracil and thymine, although intersystem crossing to the triplet manifold is observed in both nucleobases, triplet state population must compete with efficient femtosecond ($\tau_1$) and picosecond ($\tau_2$) internal conversion pathways from the $S_2$ ($^1\pi\pi^*$) and $S_1$ ($^1n\pi^*$) states, respectively, back to the ground state. This suggests that only a small fraction of the initially photoexcited $S_2$ ($^1\pi\pi^*$) state population will reach the lowest-energy triplet state in both uracil and thymine. Currently, it is unknown whether the minor fraction of the population trapped in the triplet state simply decays back to the ground state in $\tau_3$, as seems to be the case for 2-thiouracil, or if it results in some type of photodamage (*e.g.* dissociation). Nonetheless, the results presented in this work highlight the importance of both structure and environment on the ability of nucleobase analogues to dissipate potentially harmful UV radiation. Both of these factors must be taken into account in the investigation of prebiotic chemistry on early Earth and in the search for the precursors to the molecules of life.

## Acknowledgment

This work was supported by the National Science Foundation grant# NSF-CHE-1362237. The authors would also like to thank Mr. T.J. Godfrey for his assistance in the laboratory. S.M., P.M., and L.G. acknowledge the Austrian Science Fund FWF (project number P25827) and the Vienna Scientific Cluster (VSC3). M.P. and C.E.C.-H. thank the National Science Foundation for support (Grant No. CHE-1255084).

# References

1. A. I. Oparin, *The Origin of Life*, Proiskhozdenie zhizni. Izd. Moskovshii Rabochii, Moscow, 1924.
2. A. C. Rios and Y. Tor, *Isr. J. Chem.*, 2013, **53**, 469-483.
3. O. Botta and J. L. Bada, *Surv. Geophys.*, 2002, **20**, 411-467.
4. M. P. Callahan, K. E. Smith, H. J. Cleaves, J. Ruzicka, J. C. Stern, D. P. Glavin, C. H. House and J. P. Dworkin, *PNAS*, 2011, **108**, 13995-13998.
5. E. Borquez, H. J. Cleaves, A. Lazcano and S. L. Miller, *Origins Life Evol. Biospheres*, 2005, **35**, 79-90.
6. L. E. Orgel, *Critical Reviews in Biochemistry and Molecular Biology*, 2004, **39**, 99-123.
7. S. L. Miller and L. E. Orgel, *The origins of life on the earth*, Prentice-Hall, 1974.
8. H. J. Cleaves, K. E. Nelson and S. L. Miller, *Naturwissenschaften*, 2006, **93**, 228-231.
9. H. L. Barks, R. Buckley, G. A. Grieves, E. Di Mauro, N. V. Hud and T. M. Orlando, *ChemBioChem*, 2010, **11**, 1240-1243.
10. R. Saladino, G. Botta, S. Pino, G. Costanzo and E. Di Mauro, *Chemical Society Reviews*, 2012, **41**, 5526-5565.
11. W. Saenger, *Principles of nucleic acid structure*, Springer-Verlag, Berlin, 1984
12. C. S. Nabel, S. A. Manning and R. M. Kohli, *ACS Chemical Biology*, 2012, **7**, 20-30.
13. T. Carell, C. Brandmayr, A. Hienzsch, M. Müller, D. Pearson, V. Reiter, I. Thoma, P. Thumbs and M. Wagner, *Angewandte Chemie International Edition*, 2012, **51**, 7110-7131.
14. S. Zhang, J. C. Blain, D. Zielinska, S. M. Gryaznov and J. W. Szostak, *PNAS*, 2013, **110**, 17732-17737.
15. J. W. Szostak, *Journal of Systems Chemistry*, 2012, **3**, 1-14.
16. A. L. Sobolewski and W. Domcke, *Europhysics News*, 2006, **37**, 20-23.
17. C. E. Crespo-Hernández, B. Cohen, P. M. Hare and B. Kohler, *Chem. Rev.* 2004, **104**, 1997-2019.
18. C. Z. Bisgaard, H. Satzger, S. Ullrich and A. Stolow, *ChemPhysChem*, 2009, **10**, 101-110.
19. N. L. Evans and S. Ullrich, *J. Phys. Chem. A*, 2010, **114**, 11225-30.
20. K. Kleinermanns, D. Nachtigallova and M. S. de Vries, *Int. Rev. Phys. Chem.*, 2013, **32**, 308-342.
21. A. Giussani, J. Segarra-Marti, D. Roca-Sanjuan and M. Merchan, *Top. Curr. Chem.*, Springer-Verlag Berlin Heidelberg, 2013.
22. M. Barbatti, A. C. Borin and S. Ullrich, *Top. Curr. Chem.*, Springer International Publishing Switzerland, 2014.
23. R. Improta, F. Santoro, and L. Blancafort, *Chem. Rev.*, Article ASAP, doi:10.1021/acs.chemrev.5b00444
24. S. Mai, M. Richter, P. Marquetand, and L González, *Top. Curr. Chem.*, Springer Berlin-Heidelberg, 2014.
25. J. J. Serrano-Perez, R. González-Luque, M. Merchan and L. Serrano-Andres *J. Phys. Chem. B* 2007, **111**, 11880-111883.
26. S. Mai, P. Marquetand, M. Richter, J. González-Vázquez and L. González, *ChemPhysChem* 2013, **14**, 2920 - 2931
27. M. Richter, S. Mai, P. Marquetand and L. González, *Phys. Chem. Chem. Phys.* 2014, **16**, 24423-24436.
28. M. Barbatti, A. C. Borin, S. Ullrich: Photoinduced Phenomena in Nucleic Acids I, *Top. Curr. Chem.* 355, Springer 2015
29. P. M. Hare, C. E. Crespo-Hernández and B. Kohler, *PNAS*, 2007, **104**, 435-440.
30. C. T. Middleton, K. de La Harpe, C. Su, Y. K. Law, C. E. Crespo-Hernández and B. Kohler *Annu. Rev. Phys. Chem.* 2009, **60**, 217-239
31. V. Vendrell-Criado, J. A. Sáez, L.-V. V.;, M. C. Cuquerella and M. A. Miranda, *Photochem. Photobiol. Sci.*, 2013, **12**, 1460-1465.
32. M. Pollum and C. E. Crespo-Hernández, *Journal of Chemical Physics*, 2014, **140**, 071101.
33. M. Pollum, L. Martínez-Fernández and C. E. Crespo-Hernández, in *Photoinduced Phenomena in Nucleic Acids I*, eds. M. Barbatti, A. C. Borin and S. Ullrich, Springer Berlin Heidelberg, 2015, vol. 356, pp. 245-327.
34. M. Pollum, S. Jockusch and C. E. Crespo-Hernández, *Physical Chemistry Chemical Physics*, 2015, **17**, 27851-27861.
35. J. P. Gobbo and A. C. Borin, *Computational and Theoretical Chemistry* 2014, **1040-1041**, 195-201.
36. S. Mai, P. Marquetand and L. González, *J. Phys. Chem. A* 2015, **119**, 9524-9533.
37. J. Jiang, T. Zhang, J. Xue, X. Zheng, G. Cui and W. Fang *J. Chem. Phys.* 2015, **143**, 175103.
38. G. Cui and W. Fang *J. Chem. Phys.* 2015, 138, 044315.
39. M. Ruckenbauer, S. Mai, P. Marquetand and L. González, *J. Chem. Phys.*, 2016, **144**, 074303.
40. H. Yu, N. L. Evans, V. G. Stavros and S. Ullrich, *Phys. Chem. Chem. Phys.*, 2012, **14**, 6266-6272.
41. N. L. Evans, H. Yu, G. M. Roberts, V. G. Stavros and S. Ullrich, *Phys. Chem. Chem. Phys.*, 2012, **14**, 10401-9.

42. H. Yu, N. L. Evans, A. S. Chatterley, G. M. Roberts, V. G. Stavros and S. Ullrich, *J. Phys. Chem. A*, 2014, **118**, 9438-44.
43. T. J. Godfrey, H. Yu and S. Ullrich, *J. Chem. Phys.*, 2014, **141**, 044314.
44. J. D. H. Eland, *Int. J. Mass Spectrom. Ion Phys.* 1969, **2**, 471-484.
45. H. Yekeler, *J. Comput. Aided Mol. Des.* 2000, **14**, 243–250.
46. B. M. Giuliano, V. Feyer, K. C. Prince, M. Coreno, L. Evangelisti, S. Melandri, W. Caminati, *J. Phys. Chem. A*, 2010, **114**, 12725-12730.
47. C. Puzzarini, M. Biczysko, V. Barone, I. Pena, C. Cabezas, J. L. Alonso, *Phys. Chem. Chem. Phys.*, 2013, **15**, 16965-16975.
48. A. Khvorostov, L. Lapinski, H. Rostkowska and M. J. Nowak, *J. Phys. Chem. A* 2005, **109**, 770-7707.
49. V. Vaquero, M. E. Sanz, J. C. López, J. L. Alonso, *J. Phys. Chem. A*, 2007, **111**, 3443-3445.
50. J. C. López, M. Isabel Peña, M. Eugenia Sanz, J. L. Alonso, *J. Chem. Phys.*, 2007, **126**, 191103.
51. J. Finley, P. Å. Malmqvist, B. O. Roos, L. Serrano-Andrés, *Chem. Phys. Lett.* 1998, **288**, 299.
52. B. O. Roos, R. Lindh, P. Å. Malmqvist, V. Veryazov, P. O. Widmark, *J. Phys. Chem. A* 2004, **108**, 2851-2858.
53. M. J. Bearpark, M. A. Robb, H. B. Schlegel, *Chem. Phys. Lett.* 1994, **223**, 269.
54. F. Aquilante, J. Autschbach, R. K. Carlson, L. F. Chibotaru, M. G. Delcey, L. De Vico, I. Fdez. Galván, N. Ferré, L. M. Frutos, L. Gagliardi, M. Garavelli, A. Giussani, C. E. Hoyer, G. Li Manni, H. Lischka, D. Ma, P. Å. Malmqvist, T. Müller, A. Nenov, M. Olivucci, T. B. Pedersen, D. Peng, F. Plasser, B. Pritchard, M. Reiher, I. Rivalta, I. Schapiro, J. Segarra-Mart, M. Stenrup, D. G. Truhlar, L. Ungur, A. Valentini, S. Vancoillie, V. Veryazov, V. P. Vysotskiy, O. Weingart, F. Zapata, R. Lindh, *J. Comput. Chem.* 2016, **37**, 506-541.
55. L. B. Clark, G. G. Peschel and I. Tinoco, Jr., *J. Phys. Chem.*, 1965, **69**, 3615.
56. M. Barbatti, A. J. A. Aquino and H. Lischka, *Phys. Chem. Chem. Phys.*, 2010, **12**, 4959-4967.
57. M. Merchan, R. González-Luque, T. Climent, L. Serrano-Andres, E. Rodriguez, M. Reguero and D. Pelaez, *J. Phys. Chem. B*, 2006, **110**, 26471-26476.
58. S. Mai, P. Marquetand and L. González, *J. Phys. Chem. A* 2015, **119**, 9524-9533.
59. H. Kang, K. T. Lee, B. Jung, Y. J. Ko and S. K. Kim, *J. Am. Chem. Soc.*, 2002, **124**, 12958-12959.
60. S. Ullrich, T. Schultz, M. Z. Zgierski and A. Stolow, *Phys. Chem. Chem. Phys.*, 2004, **6**, 2796-2801.
61. C. Canuel, M. Mons, F. Piuzzi, B. Tardivel, I. Dimicoli and M. Elhanine, *J. Chem. Phys.*, 2005, **122**, 074316.
62. E. Samoylova, H. Lippert, S. Ullrich, I. V. Hertel, W. Radloff and T. Schultz, *J. Am. Chem. Soc.*, 2005, **127**, 1782-1786.
63. N. Gador, E. Samoylova, V. R. Smith, A. Stolow, D. M. Rayner, W. Radloff, I. V. Hertel and T. Schultz, *J. Phys. Chem. A*, 2007, **111**, 11743-11749.
64. E. Samoylova, T. Schultz, I. V. Hertel and W. Radloff, *Chem. Phys.*, 2008, **347**, 376-382.
65. J. González-Vazquez, L. González, E. Samoylova and T. Schultz, *Phys. Chem. Chem. Phys.*, 2009, **11**, 3927-3934.
66. M. Kotur, T. C. Weinacht, C. Zhou and S. Matsika, *IEEE Journal of Selected Topics in Quantum Electronics*, 2012, **18**, 187-194.
67. M. Ligare, F. Siouri, O. Bludsky, D. Nachtigallová and M. S. de Vries, *Phys. Chem. Chem. Phys.*, 2015, **17**, 24336-24341.
68. M. M. Brister and C. E. Crespo-Hernández, *J. Phys. Chem. Lett.*, 2015, **6**, 4404-4409.
69. D. Asturiol, B. Lasorne, M. A. Robb and L. Blancafort, *J. Phys. Chem. A* 2009, **113**, 10211-10218.
70. H. R. Hudock, B. G. Levine, A. L. Thompson, H. Satzger, D. Townsend, N. Gador, S. Ullrich, A. Stolow, and T. J. Martinez, *J. Phys. Chem. A*, 2007, **111**, 8500-8508
71. J. J. Szymczak, M. Barbatti, J. T. Soo Hoo, J. A. Adkins, T. L. Windus, D. Nachtigallová and H. Lischka, *J. Phys. Chem. A*, 2009, **113**, 12686–12693
72. D. Nachtigallová, A. J. A. Aquino, J. J. Szymczak, M. Barbatti, P. Hobza, and H. Lischka, *J. Phys. Chem. A*, 2011, **115**, 5247–5255
73. M. Barbatti, A. J. A. Aquino, J. J. Szymczak, D. Nachtigallova, P. Hobza and H. Lischka, *PNAS*, 2010, **107**, 21453-21458.
74. B. P. Fingerhut, K. E. Dorfman, and S. Mukamel, *J. Chem. Theory Comput.*, 2014, **10**, 1172–1188
75. A. Nakayama, G. Arai, S. Yamazaki, and T. Taketsugu, *J. Chem. Phys.*, 2013, **139**, 214304.
76. S. Yamazaki and T. Taketsugu, *J. Phys. Chem. A*, 2011, **116**, 491-503.
77. S. Matsika, M. Spanner, M. Kotur and T. C. Weinacht, *J. Phys. Chem. A*, 2013, **117**, 12796-12801.

78. B. B. Brady, L. A. Peteanu and D. H. Levy, *Chem. Phys. Lett.*, 1988, **147**, 538-543.
79. K-W. Choi, J-H. Lee and S. K. Kim, *Chem. Commun.*, 2006, **1**, 78-79.
80. D. Dougherty, K. Wittel, J. Meeks and S. P. McGlynn, *Am. Chem. Soc.*, 1976, **98**, 3815.
81. B. Lasorne, M. A. Robb and L. Blancafort, *J. Phys. Chem. A* 2009, **113**, 10211-10218.
82. S. Perun and A. L. Sobolewski, *J. Phys. Chem. A*, 2006, **110**, 13238-13244.
83. M. Etinski, T. Fleig and C. M. Marian, *J. Phys. Chem. A*, 2009, **113**, 11809-11816.
84. H.-W. Jochims, M. Schwell, H. Baumgartel and S. Leach, *Chem. Phys.*, 2005, **314**, 263–282.
85. J. J. Serrano-Perez, R. González-Luque, M. Merchan and L. Serrano-Andres, *J. Phys. Chem. B*, 2007, **111**, 11880-11883.
86. N. J. Kim, J. Chang, H. M. Kim, H. Kang, T. K. Ahn, J. Heo and S. K. Kim, *Chem. Phys. Chem.*, 2011, **12**, 1935-1939.
87. A. R. Katritzky, M. Szafran and G. Pfister-Guillouzo, *J. Chem. Soc. Perkin Trans.* 1990, **2**, 871-876.
88. C. Salet and R. Bensasson, *Photochem. Photobiol.*, 1975, **22**, 231-235.
89. S. J. Milder, D. S. Kliger, *J. Am. Chem. Soc.*, 1985, **107**, 7365–7373